\documentclass[aps,pra, twocolumn,a4paper,showpacs]{revtex4}
\usepackage{amsmath}
\usepackage{bm}
\usepackage{graphicx}
 \usepackage{epsfig}

 \def\BEA {\begin{eqnarray}}
 \def\EEA {\end{eqnarray}}
 \def\BE {\begin{equation}}
 \def\EE {\end{equation}}
 \def\BA {\begin{array}}
 \def\EA {\end{array}}
 \def\NN {\nonumber}
  \def\re {{\rm{Re} \,}}
 \def\im {{\rm{Im} \,}}
 
\begin{document}

 \title{Spatial multipartite entanglement and localization of entanglement}

 \author{D.~Daems and N.~J.~Cerf}
 \affiliation{QuIC, Ecole Polytechnique, Universit\'e Libre de Bruxelles,
1050 Brussels, Belgium}

 \begin{abstract}
We present a simple model together with its physical implementation which allows one to generate
multipartite entanglement between several spatial modes of the electromagnetic field.
It is based on  parametric down-conversion with $N$ pairs of symmetrically-tilted plane waves serving as a pump. 
The characteristics of this spatial entanglement are investigated in the cases of zero as well as nonzero phase mismatch. Furthermore, the phenomenon of entanglement localization in just two spatial modes is studied in detail and results in an enhancement of the entanglement by a factor $\sqrt{N}$.
 \end{abstract}
 \pacs{03.67.Bg, 42.65.Lm}
 \maketitle

\section{Introduction}
The existence of entanglement  between more than two parties for discrete or continuous quantum variables is one of the most striking
predictions of quantum mechanics. The interest in multipartite entanglement is
motivated not only by fundamental questions but also by its potential for
applications in quantum communication technologies.
Possible applications of continuous-variable (CV) multipartite entanglement include quantum teleportation networks, quantum telecloning, controlled
quantum dense coding, and quantum secret sharing (see
Ref.~\cite{Peng2007} and references therein).

Characterizing multipartite entanglement has raised much interest since the pioneering work of 
Coffman, Kundu and Wootters \cite{Coffman2000}.
They have established for a three-qubit system and
conjectured for $N$-qubit systems the so-called monogamy of quantum
entanglement, constraining the maximum entanglement between partitions of
a multiparty system. More recently, Adesso, Serafini and Illuminati \cite{ASI04}-\cite{Cerf2007} have introduced the
continuous-variable tangle as a measure of multipartite entanglement for
continuous-variable multimode Gaussian states. In particular, they have demonstrated
that this tangle satisfies the Coffman-Kundu-Wootters monogamy inequality.
The conjecture of Ref. \cite{Coffman2000} has been proven by Osborne and Verstraete \cite{OV2006}. The corresponding proof for Gaussian states has been obtained by Hiroshima, Adesso and Illuminati \cite{HAI2007}.

A number of schemes to generate CV multipartite entanglement have
been proposed theoretically and realized experimentally
in recent years.  There were first the passive optical
scheme using squeezed states mixed with beam
splitters~\cite{vanLoock2000}. Then came the active schemes in which
multipartite entanglement is created as a result of parametric interaction
of several optical waves such as
cascaded/concurrent~\cite{Pfister2004,Guo2005,Yu2006},
interlinked~\cite{Ferraro2004,Olsen2006}, or consecutive parametric
interactions~\cite{Rodionov2004}.
These  schemes generally neglect
the spatial structure of the electromagnetic field. 

It is natural to investigate whether the spatial modes
can also serve for the creation of CV multipartite entanglement.
This question has been adressed recently in Ref.   \cite{DBCK10}  where  a simple active scheme was proposed for the 
creation of tripartite entanglement between spatial modes of the
electromagnetic field in the process of parametric interaction.
It consists in pumping a
nonlinear parametric medium by a coherent combination of several tilted
plane monochromatic waves which is called a spatially-structured pump. Since
the pump photon can be extracted from any of these waves and the pair of
down-converted photons emitted in different directions according to the
phase-matching condition, that scheme allows for the creation of tripartite
entanglement between spatial modes of the down-converted field. 
The aim of this paper is to generalize the results of Ref. \cite{DBCK10} to the generation of spatial multipartite entanglement.
For that purpose, we present a scheme with an arbitrary number $2N$ of tilted plane waves pumping a parametric medium. 
An interesting feature of this realistic proposal is  the
possibility of localizing the created spatial multipartite entanglement
in just two well-defined spatial modes formed as a linear combination of all
the modes participating in the down-conversion process. 
The
possibility of entanglement localization was introduced by 
Serafini, Adesso and Illuminati \cite{ASI05} and a  physical implementation was also given in terms of 2N-1 beam splitters and 2N single-mode squeezed inputs based on \cite{VF03}.

The paper is organized as follows. Section~II is devoted to spatial multipartite entanglement. We present the process of parametric down-conversion with a spatially structured pump consisting of $N$ pairs of symmetrically-tilted
plane waves in Subsec.~II~A. The evolution equations for the field operators are given and solved in the rotating wave approximation for possibly non-zero constant phase mismatch. The genuine multipartite entanglement is then studied In Subsec.~II~B. Section III is dedicated to the phenomenon of entanglement localization. It is presented explicitly in Subsec.~III~A and a quantitative characterization of the spatial entanglement generated by this process is given in Subsec.~III~B. The scaling with  $N$ as well as the dependence on the phase mismatch are explicitly provided. Conclusions are drawn in Section~IV.

\section{Spatial multipartite entanglement}
\subsection{Parametric down-conversion with spatially-structured pump}
We consider a system of pumps which consists of $N$ pairs of symmetrically tilted plane waves:
\begin{equation}
E_p({\bf r}) = \frac{\alpha}{4\pi^2} \sum_{d=1}^N \left(e^{i{\bf q}_{d}.{\bf r}}+e^{-i{ \bf q}_{d}.{\bf r}}\right),
\end{equation} 
with  $\bf{r}=(x,y)$, the vector of coordinates in the plane of the crystal entering face, and $\bf{q}=(k_x,k_y)$, the projection of the three-dimensional wave vector ${\bf k}$ in that plane.
Its Fourier transform  in the $xy$ plane of the (infinite) crystal (entering face) corresponds to Dirac delta's centered in ${\bf q_d}$ et ${\bf -q_d}$ 
\begin{equation}
E_p(\bf{q}) = \alpha \sum_{d=1}^N \left[\delta({\bf q}-{\bf q}_{d}) + \delta({\bf q}+{\bf q}_{d}) \right].
\label{TFN}
\end{equation}
This scheme is illustrated on Fig. 1 for  $N=2$ and 4. The little circles represent the projections of the pump wave vectors in the $xy$ plane of the crystal entering face.
 The plane depicted in Fig. 2 is the one going through one of the $N$  pairs of opposed pumps  on  Fig. 1, for example the $yz$ plane.

 \begin{center}
\begin{figure}
\includegraphics[scale=0.45]{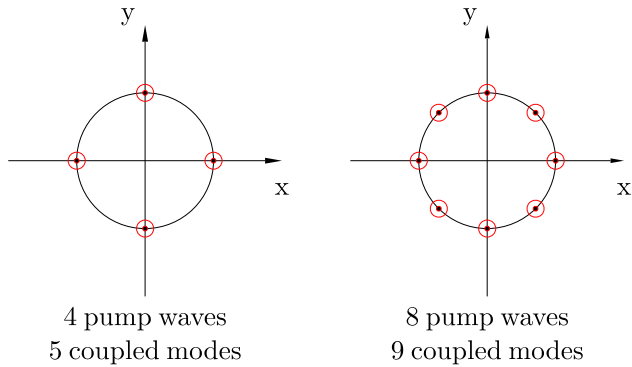}
\caption{(Color online)  Scheme for the generation  of spatial multipartite entanglement with $2N$ symmetrically tilted pump waves (illustrated for $N=2$ and 4). The little circles represent the projections of the pump wave vectors in the $xy$ plane of the crystal entering face.
} 
\end{figure}
\end{center}

 \begin{center}
\begin{figure}
\includegraphics[scale=0.45]{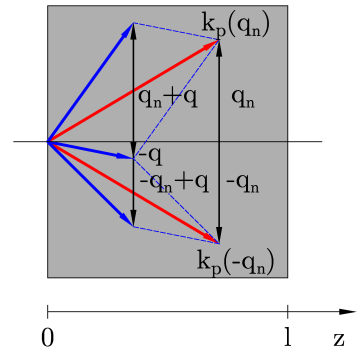}
\caption{(Color online) The plane depicted is the one passing through one of the $N$  pairs of opposed pumps on Fig. 1 (e. g. the $yz$ plane).
The tilted pumps have wave vectors ${\bf k}_p (\pm {\bf q}_n)$ with $n$ ranging from $1$ to $N$. The transverse (vertical) components are $\pm {\bf q}_n$.
} 
\end{figure}
\end{center}
  
The evolution of the creation and annihilation operators, $\hat{a}$ and $\hat{a}^\dag$, associated to the propagation and diffraction of the quantized electromagnetic field through the nonlinear parametric medium is described by  \cite{Kolobov1999}
\begin{equation}
\frac{\partial}{\partial z}\hat{a}(z,{\bf q}) = \lambda \int {\bf d q^\prime} E_p({\bf q}-{\bf q^\prime}) \hat{a}^\dag(z,-{\bf q^\prime}) e^{i\Delta({\bf q},-{\bf q^\prime})z}.
\label{evol4}
\end{equation} 
Here $z$ is the longitudinal coordinate and $\Delta$ is the phase mismatch defined as
\begin{equation}
\Delta({\bf q},-{\bf q^\prime}) = k_z({\bf q})+k_z(-{\bf q^\prime})-k_{pz}({\bf q}-{\bf q^\prime}),
\end{equation}
where  $k_z({q})$ and $k_z(-{q})$ are the longitudinal component of the two incoming wave vectors and $k_{pz}$ is the longitudinal component of the pump wave vector.
The conservation of energy and momentum imply that we have to consider a set of $2N+1$ coupled equations, obtained upon substitution of ({\ref{TFN}) into (\ref{evol4}), for the waves exiting the crystal with wave vector transverse components  ${\bf q}$, ${\bf q}_d-{\bf q}$ and $-{\bf q}_d-{\bf q}$ with $d=1,\cdots,N$
\begin{widetext}
\BEA
\frac{\partial}{\partial z}\hat{a}(z,{\bf q}) & = & \alpha \lambda \sum_{d=1}^N \left\{\hat{a}^\dag(z,{\bf q}_d-{\bf q})e^{i\Delta({\bf q},{\bf q}_d-{\bf q})z}+\hat{a}^\dag(z,-{\bf q}_d-{\bf q})e^{i\Delta({\bf q},-{\bf q}_d-{\bf q})z} \right\} \label{NeqFULL} \\
\frac{\partial}{\partial z}\hat{a}(z,\pm {\bf q}_j+{\bf q}) & = & \ \alpha \lambda \sum_{d=1}^N \left\{\hat{a}^\dag(z,{\bf q}_d  \mp {\bf q_j}-{\bf q})e^{i\Delta({\bf q},{\bf q}_d  \mp {\bf q_j}-{\bf q})z}+
\hat{a}^\dag(z,-{\bf q}_d  \mp {\bf q_j}-{\bf q})e^{i\Delta({\bf q},-{\bf q}_d  \mp {\bf q_j}-{\bf q})z}\right\}  \ . \NN
\EEA
\end{widetext}
In the second equation, the phase mismatch is higher for the contributions with $d \neq j$ which shall therefore be neglected.
This corresponds to the usual rotating wave approximation.
The other phase mismatches are all taken equal to $\Delta$ which amounts to imposing some symetries on $\Delta(\pm {\bf q}_d\pm q,\pm {\bf q})$.
One may introduce a  renormalized phase mismatch $\delta$ and a new relevant variable $\tilde r\equiv \sqrt{2} \alpha \lambda z$ which combines the interaction strength $\alpha \lambda$ and the longitudinal coordinate $z$:
\BEA
\delta&\equiv& \frac{\Delta}{2 \sqrt{2} \alpha \lambda} \NN\\
\tilde r&\equiv& \sqrt{2} \alpha \lambda z.
\EEA
We shall use the following notation
\BEA
a_0(\tilde r)&\equiv&\hat{a}(z,{\bf q})\\
a_{n_\pm}(\tilde r)&\equiv& \hat{a}(z,\pm {\bf q}_n+{\bf q}) \qquad n=1,\cdots,N,\NN
\EEA
and shall consider only the zeroth spatial Fourier components of the field, ${\bf q}=0$. Physically, this corresponds to photodetection of the light field by a single large photodetector without spatial resolution. These modes are depicted in  Fig.~2: the long (blue) arrows pertains to $\hat{a}_{n_\pm}(0)$, the short one to $\hat{a}_0^\dag(0)$.
In this setting, one can then rewrite (\ref{NeqFULL})  as
\begin{eqnarray}
\frac{d}{d\tilde r}\hat{a}_0 & = & \frac{e^{2i\delta \tilde r}}{\sqrt{2}}    \sum_{d=1}^N \left( \hat{a}_{d_+}^\dag +  \hat{a}_{d_-}^\dag \right) \nonumber\\
\frac{d}{d\tilde r}\hat{a}_{n_\pm} & = &  \frac{e^{2i\delta \tilde r} }{\sqrt{2}}   \hat{a}_0^\dag  \qquad n=1,\cdots,N.
\label{3eq}
\end{eqnarray}
We may solve this set of equations and obtain for the fields at the output of the crystal, $r=\tilde r|_{z=\ell}$:
\begin{eqnarray}\label{solneq}
\hat{a}_0(r) &=& U(r)\hat{a}_0(0) + \frac{V(r) }{\sqrt{2N}}    \sum_{d=1}^N \left\{ \hat{a}_{d_+}^\dag(0) +  \hat{a}_{d_-}^\dag(0) \right\}\NN \\
\hat{a}_{n_\pm}(r) &=&\hat{a}_{n_\pm}(0)+ \frac{U(r)-1}{\sqrt{2N}} \sum_{d=1}^N \left\{ \hat{a}_{d_+}(0) +  \hat{a}_{d_-}(0) \right\} \NN\\
&+& \frac{V(r)}{\sqrt{2N}}\hat{a}_0^\dag(0) 
  \qquad n=1,\cdots,N.
\end{eqnarray}
The functions $U(r)$ and $V(r)$, which satisfy $|U(r)|^2-|V(r)|^2=1$, are given by
\begin{eqnarray}\label{UV}
U(r) & = & e^{i \delta r} \left( \cosh(\sqrt{N}\gamma r) - i\frac{\delta}{\sqrt{N}\gamma}\sinh(\sqrt{N}\gamma r) \right) \nonumber\\
V(r) & = & e^{i \delta r}  \frac{1}{\gamma}   \sinh(\sqrt{N}\gamma r) ,
\end{eqnarray}
where 
$\gamma$ depends on the reduced phase mismatch $\delta$,
\begin{equation}
\gamma = \sqrt{1-\frac{\delta^2}{N}}.
\label{Gam}
\end{equation} 
In the next subsection we shall characterize these solutions for the fields at the output of the crystal.

\subsection{Genuine multipartite entanglement}
The parametric down-conversion process preserves the Gaussian character of incoming modes. Hence, the outgoing modes are Gaussian states which are thus completely characterized by the covariance matrix associated to their quadrature components:
 \BEA
 \hat{x}_0(r)& \equiv& 2 {\rm Re}\, \hat{a}_0(r)\NN\\
 \hat{p}_0(r) &\equiv& 2  {\rm Im}\,  \hat{a}_0(r)\NN\\
  \hat{x}_{n_\pm}(r)& \equiv& 2 {\rm Re}\, \hat{a}_{n_\pm}(r) \qquad  n=1,\cdots,N \NN\\ 
 \hat{p}_{n_\pm}(r) &\equiv& 2  {\rm Im}\,  \hat{a}_{n_\pm}(r).
 \EEA
From the solution ({\ref{solneq}) one obtains
\begin{widetext}
\begin{eqnarray}\label{6quad}
\hat{x}_0(r) & = &\re U(r) \hat{x}_0(0) - \im U(r) \hat{p}_0(0)+  \sum_{d=1}^N \left\{ \re V(r)  \frac{ \hat{x}_{d_+}(0) +  \hat{x}_{d_-}(0)} {\sqrt{2N}}  + \im V(r) \frac{\hat{p}_{d_+}(0) +  \hat{p}_{d_-}(0)}{\sqrt{2N}} \right\}\nonumber\\
\hat{p}_0(r) & = &\im U(r) \hat{x}_0(0)+ \re U(r) \hat{p}_0(0)  +  \sum_{d=1}^N \left\{ \im V(r) \frac{\hat{x}_{d_+}(0) +  \hat{x}_{d_-}(0)}{\sqrt{2N}}  - \re V(r)  \frac{ \hat{p}_{d_+}(0) +  \hat{p}_{d_-}(0)} {\sqrt{2N}}   \right\}\nonumber\\
\hat{x}_{n_\pm}(r) & = & \frac{\re V(r)}{\sqrt{2N}} \hat{x}_0(0) +\frac{\im V(r)}{\sqrt{2N}} \hat{p}_0(0)   + \hat{x}_{n_\pm}(0)  +  \sum_{d=1}^N \left\{   (\re U(r) -1) \frac{\hat{x}_{d_+}(0) +  \hat{x}_{d_-}(0) }{2N}  - \im U(r)\frac{\hat{p}_{d_+}(0) +  \hat{p}_{d_-}(0) }{2N}  \right\}     \nonumber\\
\hat{p}_{n_\pm}(r) & = &  \frac{\im V(r) }{\sqrt{2N}} \hat{x}_0(0) - \frac{\re V(r)  }{\sqrt{2N}}  \hat{p}_0(0)+ \hat{p}_{n_\pm}(0)  +  \sum_{d=1}^N \left\{  \im U(r) \frac{\hat{x}_{d_+}(0) +  \hat{x}_{d_-}(0) }{2N}  +  (\re U(r) -1) \frac{\hat{p}_{d_+}(0) +  \hat{p}_{d_-}(0) }{2N}  \right\}  .   \nonumber\\
\label{quad}
\end{eqnarray}
\end{widetext}
We can now determine the covariance matrix elements $\sigma_{ij} \equiv  \left\langle \left(\Delta\hat{\xi}_i \Delta\hat{\xi}_j + \Delta\hat{\xi}_j \Delta\hat{\xi}_i\right)/2 \right\rangle $ of the output state $\rho$ with $ \left\langle . \right\rangle \equiv \textrm{Tr}\left[{\rho} .  \right] $ and 
$\Delta\hat{\xi}_i \equiv  \hat{\xi}_i-\left\langle\hat{\xi}_i\right\rangle$ where $\hat{\xi}_i$ is some component of the vector $\hat{{\bf \xi}} = \left(\hat{x}_0,\hat{p}_0,\hat{x}_{1_+},\hat{p}_{1_+},\hat{x}_{1_-},\hat{p}_{1_-},\hat{x}_{2_+},\hat{p}_{2_+},\cdots, \hat{x}_{N_-},\hat{p}_{N_-}\right)$.
Taking into account that all inputs are vacuum states we obtain from (\ref{quad}) 
\begin{eqnarray}\label{paramCM3}
\!\!\!&\!\!\!&\!\!\! \left\langle [\hat{x}_0(r)]^2 \right\rangle= \left\langle [\hat{p}_0(r)]^2 \right\rangle= a\NN\\
\!\!\!&\!\!\!&\!\!\! \left\langle [\hat{x}_{n_\pm}(r)]^2 \right\rangle=\left\langle [\hat{p}_{n_\pm}(r)]^2\right\rangle=1+\frac{a-1}{2N}\NN\\
\!\!\!&\!\!\!&\!\!\!  \left\langle \hat{x}_{n_\pm} (r) \hat{x}_{d_\pm}(r)\right\rangle=\left\langle \hat{x}_{n_\pm} (r) \hat{x}_{d_\mp}(r)\right\rangle=\frac{a-1}{2N}\NN\\
\!\!\!&\!\!\!&\!\!\!  \left\langle \hat{p}_{n_\pm} (r) \hat{p}_{d_\pm}(r)\right\rangle=\left\langle \hat{p}_{n_\pm} (r) \hat{p}_{d_\mp}(r)\right\rangle=\frac{a-1}{2N}\NN\\
\!\!\!&\!\!\!&\!\!\!  \left\langle \hat{x}_0(r) \hat{x}_{n_\pm}(r)\right\rangle = -\left\langle \hat{p}_0 (r)\hat{p}_{n_\pm}(r)\right\rangle =   \frac{b}{\sqrt{2N}}\NN \\ 
\!\!\!&\!\!\!&\!\!\!  \left\langle \hat{x}_0(r) \hat{p}_{n_\pm}(r)\right\rangle = \left\langle \hat{p}_0 (r)\hat{x}_{n_\pm}(r)\right\rangle = \frac{c}{\sqrt{2N}}\NN\\
\!\!\!&\!\!\!&\!\!\!  \left\langle \hat{x}_{0} (r) \hat{p}_{0}(r)\right\rangle=\left\langle \hat{x}_{n_\pm}  (r)\hat{p}_{d_\pm}(r)\right\rangle=\left\langle \hat{x}_{n_\pm} (r) \hat{p}_{d_\mp}(r)\right\rangle=0,  \NN\\
\label{cov}
\end{eqnarray}   
for $ n,d=1,\cdots,N,$ with
\begin{eqnarray}\label{abc}
&&a \equiv  |U|^2+|V|^2  \NN \\
&&b  \equiv   2({\rm Re}\, U \ {\rm Re}\, V -  {\rm Im}\,  U  \ {\rm Im}\,  V)  \NN \\ 
&& c  \equiv 2({\rm Re}\, U \ {\rm Im}\,  V +  {\rm Im}\,  U {\rm Re}\, V)  .\end{eqnarray}

Ordering the lines and columns according to $\hat{x}_0$, $\hat{p}_0$, $\hat{x}_{1_+}$, $\hat{p}_{1_+}$, $\hat{x}_{1_-}$, $\hat{p}_{1_-}$, $\hat{x}_{2_+}$, $\hat{p}_{2_+}$, 
$\cdots$  $\hat{x}_{N_-}$, $\hat{p}_{N_-}$, the covariance matrix reads then
\begin{equation}
\sigma = 
\left(\begin{array}{cccccccc}
A & D & D & D & \cdots & D &D & D \\
D & B & C & C & \cdots &  C& C & C\\
D & C & B& C & \cdots & C & C & C \\
D & C & C& B & \cdots & C & C & C \\
\vdots & \vdots & \vdots & \vdots & \ddots & \vdots & \vdots & \vdots \\
D & C & C & C & \cdots & B &  C & C\\
D & C & C & C & \cdots &  C & B & C\\
D & C & C & C & \cdots &  C &C & B\\
\end{array}\right) ,
\label{CM3}
\end{equation} 
where $A$, $B$, $C$ and $D$ are the following $2 \times 2$ matrices
\BEA
&&A=a I_2, \qquad B=1 +\frac{a-1}{2N} I_2, \qquad C= \frac{a-1}{2N} I_2\NN\\
&&D=\frac{1}{\sqrt{2N}} \left(\begin{array}{cr}
b & c \\
c & -b
\end{array}\right), \qquad  I_2= \left(\begin{array}{cc}
1 & 0 \\
0 & 1
\end{array}\right).
\EEA
Its symplectic eigenvalues $\{\nu_i\}$ are $\sqrt{a^2-b^2-c^2}$ (with a fourfold degeneracy) and $1$ (with a $2N-3$ degeneracy). Noting that $b^2+c^2=4|U|^2 |V|^2$ implies that $\sqrt{a^2-b^2-c^2}=1$.
The product of symplectic eigenvalues is thus unity and, as expected for a unitary evolution, the output state remains pure.

The covariance matrix (\ref{CM3}) is bisymmetric, i. e.  it is invariant under the permutation of  quadratures 
$\hat{x}_{n_\pm}(r),\hat{p}_{n_\pm}(r) \leftrightarrow \hat{x}_{d_\pm}(r),\hat{p}_{d_\pm}(\ell)$ or $\hat{x}_{d_\mp}(r),\hat{p}_{d_\mp}(\ell)$. As a consequence, it has the multipartite entanglement structure of covariance matrices associated to bisymmetric (1+2N)-mode Gaussian states considered in  Ref. \cite{ASI05}. We now show that the output state obtained here exhibits genuine multipartite entanglement in the sense of Ref. \cite{VF03}. For that purpose, we have to verify that the following condition on covariance matrix elements is violated
 \BEA
Q&\equiv &\left\langle\left( \hat{x}_0(r)- \frac{1}{\sqrt{ 2N}}\sum_{n=1}^N\left\{ \hat{x}_{n_+}(r)+\hat{x}_{n_-}(r)\right\}\right)^2 \right\rangle \NN\\
&+&  \left\langle  \left( \hat{p}_0(r)+ \frac{1}{\sqrt{ 2N}}\sum_{n=1}^N\left\{ \hat{p}_{n_+}(r)+\hat{p}_{n_-}(r)\right\}\right)^2  \right\rangle \geq \frac{1}{2N}.\NN\\ \label{C}
 \EEA
  From (\ref{cov})-(\ref{abc}) one deduces that 
  \BEA
  Q&=&4 (a-b)\\
  &=& 4\left( [\re U(r) -\re V(r)]^2 +  [\im U(r) + \im V(r)]^2  \right) .  \label{Q}\NN
  \EEA
  When the phase matching condition is satisfied, $\delta=0$, one deduces from (\ref{UV}) that $U(r)=\cosh \sqrt{N}r$ and $V(r)=\sinh \sqrt{N}r$. It follows that (\ref{Q}) reduces to
  \BE
  Q=4 \{\cosh (\sqrt{N}r)-\sinh(\sqrt{N}r)\}=4 e^{-2\sqrt{N}r}.
  \EE
This quantity is smaller than $1/2N$  if the squeezing parameter $r$ satisfies
\BE
 r= \sqrt{2}\alpha \lambda \ell > \frac{3 \ln (2N)}{2 \sqrt{N}}.
 \EE
 Under this condition on  the pump amplitude $\alpha$, the coupling parameter $\lambda$ and the crystal length $\ell$, the output state $\rho$ produced by the above parametric down-conversion process with $2N$ symmetrically-tilted
plane waves therefore exhibits genuine multipartite entanglement.
 This generalizes the result obtained in Ref. \cite{DBCK10} in the case of tripartite entanglement ($N=1$).
  
  For a small nonzero phase mismatch, we can expand (\ref{Q}) around $\delta=0$, which yields
  \BEA
  Q&=&4 e^{-2\sqrt{N}r} +\frac{\delta^2}{N}\left(  [3-4Nr^2] e^{-2\sqrt{N}r}   \right.  \NN\\
  &+& \left. 4[ 2 \sqrt{N} r-1]  +[2 \sqrt{N} r-1]^2 e^{2\sqrt{N} r} \right) + O(\delta^4).\NN\\
  \EEA
Owing to the last term, this quantity increases exponentially with $\sqrt{N}r$.
As a consequence, $Q$ remains smaller than $1/2N$ only for low values of the phase mismatch.
This suggests that the genuine multipartite entanglement, at least when it is estimated with the criterion  (\ref{C}) is very sensitive to $\delta$. This situation is in contrast with the phenomenon that we study in the next section, namely the localization of entanglement which will be shown to be robust with respect to the phase mismatch.

\section{Localization of entanglement}
\subsection{Beamsplitting the output state}
It has been shown \cite{ASI05} that the entanglement of bisymmetric $(m+n)$-mode Gaussian states is unitarily {\it localizable}, i. e., that, through local unitary operations, it may be fully concentrated in a single pair of modes. We shall study explicitly this phenomenon here.
For that purpose we shall perform the following unitary transformation based on discrete Fourier series,
\begin{eqnarray}\label{unitary}
\hat{a}'_0(\tilde r)&=& \hat{a}_0(\tilde r)  \NN\\
\hat{a}'_k(\tilde r)  &=&\frac{1}{\sqrt{2N}}  \sum_{n=1}^N   e^{-\pi i (k-1) \frac{n-1}{N}}    \{      \hat{a}_{n_+} (\tilde r)  -e^{-\pi i k}   \hat{a}_{n_-}(\tilde r)  \}\NN\\
& & k=1,\cdots,2 N. \
\end{eqnarray}
Physically, this corresponds to beamsplitting the quantized fields at the output of the crystal.
As a consequence, from (\ref{quad}) one deduces that the new fields at the output of the crystal are transformed to the following ones
\begin{eqnarray}\label{sol3eq}
\hat{a}'_0(r) & = & U(r)\hat{a}'_0(0) + V(r){\hat{a}'_1}{^\dag}(0)\NN \\
\hat{a}'_1(r) & = & U(r)\hat{a}'_1(0) + V(r){\hat{a}'_0}{^\dag}(0) \nonumber\\
\hat{a}'_k(r) & = &\hat{a}'_k(0)  \qquad k=2,\cdots, 2 N,
\end{eqnarray}
 Accordingly, the quadratures are now given by
\begin{widetext}
\begin{eqnarray}\label{6quad2}
\hat{x}'_0(r) & = &{\rm Re} \,U(r) \hat{x}'_0(0)- {\rm Im}\,  U(r) \hat{p}'_0(0)+{\rm Re} \,V(r) \hat{x}'_1(0)+ {\rm Im}\,  V(r) \hat{p}'_1(0)\nonumber\\
\hat{p}'_0(r) & = & {\rm Re}\, U(r) \hat{p}'_0(0)+ {\rm Im}\,  U(r) \hat{x}'_0(0)-{\rm Re}\, V(r) \hat{p}'_1(0)+ {\rm Im}\,  V(r) \hat{x}'_1(0)\nonumber\\
\hat{x}'_1(r) & = & {\rm Re}\, U(r) \hat{x}'_1(0)- {\rm Im}\,  U(r) \hat{p}'_1(0)+{\rm Re}\, V(r) \hat{x}'_0(0)+ {\rm Im}\,  V(r) \hat{p}'_0(0) \nonumber\\
\hat{p}'_1(r) & = & {\rm Re}\, U(r) \hat{p}'_1(0)+ {\rm Im}\,  U(r) \hat{x}'_1(0)-{\rm Re}\, V(r) \hat{p}'_0(0)+ {\rm Im}\,  V(r) \hat{x}'_0(0)\\
\hat{x}'_k(r) & = & \hat{x}'_k(0) \qquad k=2,\cdots,2N\nonumber\\
\hat{p}'_k(r) & = & \hat{p}'_k(0)\nonumber.
\end{eqnarray}
\end{widetext}
Hence, the covariance matrix elements associated to the density matrix $\rho'$ are
\begin{eqnarray}\label{paramCM32}
&&\left\langle[ \hat{x}'_0(r)]^2\right\rangle= \left\langle[ \hat{p}'_0(r)]^2\right\rangle=\left\langle [\hat{x}'_1(r)]^2\right\rangle=\left\langle [\hat{p}'_1(r)]^2\right\rangle= a\NN \\
&&\left\langle \hat{x}'_0(r) \hat{x}'_1(r)\right\rangle = -\left\langle \hat{p}'_0(r) \hat{p}'_1(r)\right\rangle =   b\NN \\ 
&& \left\langle \hat{x}'_0(r) \hat{p}'_1(r)\right\rangle = \left\langle \hat{p}'_0(r) \hat{x}'_1(r)\right\rangle = c\NN\\
&& \left\langle [\hat{x}'_k(r)]^2\right\rangle= \left\langle [\hat{p}'_k(r)]^2\right\rangle=1 \qquad k=2,\cdots,2 N,
\end{eqnarray}  
and the other ones are zero.
Ordering the lines and columns according to $\hat{x}'_0$, $\hat{p}'_0$, $\hat{x}'_1$, $\hat{p}'_1$, $\hat{x}'_2$, $\hat{p}'_2$, 
$\cdots$  $\hat{x}'_{2N}$, $\hat{p}'_{2N}$, the covariance matrix reads
\begin{equation}
\sigma' = 
\left(\begin{array}{cccc}
a & 0 & b & c \\
0 & a & c & -b\\
b & c & a & 0 \\
c & -b & 0 & a \\
\end{array}\right) \bigoplus I_{2(2N-1)},
\label{CM4}
\end{equation} 
where $I_{k}$ is the $k\times k$ unity matrix.
Its symplectic eigenvalues $\{\nu'_i\}$ are the same as those of $\sigma$ as $\sigma'$ is obtained by congruence,
\BE
\sigma'=S^{\rm T} \sigma S,
\EE
where the elements of the symplectic transformation $S$ are given by (\ref{unitary}).
In the next subsection, we investigate the non-separability properties of the pertaining state $\rho'$.

\subsection{Logarithmic negativity}
The covariance matrix (\ref{CM4}) is bisymmetric, i. e.  the local exchange of any pairs of modes within its two diagonal blocks leaves the matrix invariant \cite{ASI05}. Notice that here the unitary transformation (\ref{unitary}) is such that the covariant matrix $\sigma'$ is block-diagonal, and invariant under the exchange of modes $0$ and $1$.

The separability of the state $\rho'$ can be determined as in Ref. \cite{ASI05} by (i) considering a partition of the system in two subsystems and (ii) investigating the positivity of the partially transposed matrix $\tilde \rho'$ obtained upon transposing the variables of only one of the two subsystems.
The positivity of partial transposition (PPT)  is a necessary condition for the separability of any bipartite quantum state \cite{P96},\cite{H96}.
It is also sufficient for the separability of $(1+n)-$mode Gaussian states \cite{S00},\cite{WW01}. 
The covariance matrix $\tilde \sigma'$ of the partially transposed state  $\tilde \rho'$  with respect to one subsystem is obtained \cite{S00}  by changing the signs of the quadratures  $p'_j$ belonging to that subsystem.

Here, because of the block-diagonal structure of $\sigma'$ we can restrict the analysis to the first block and consider the transposition of mode 1, i.e. change the sign of $p'_1$:
\begin{equation}
\tilde \sigma' = 
\left(\begin{array}{cccc}
a & 0 & b & c \\
0 & a & -c & b\\
b & -c & a & 0 \\
c & b & 0 & a \\
\end{array}\right) \bigoplus I_{2(2N-1)},
\label{TCM3}
\end{equation} 
Its symplectic eigenvalues $\{\tilde \nu'_j\}$ are $1$ and
\BE
\tilde \nu'_\pm \equiv a \pm  \sqrt{b^2+c^2}=(|U| \pm |V|)^2. \label{nu}
\EE
The necessary and sufficient PPT condition for the separability of the state $\rho'$ amounts to having  $\tilde \nu'_j \geq 1 \forall j$.  We can therefore focus on the smallest eigenvalue $\tilde \nu'_-$.
The extent to which this criterion is violated is measured by $\mathcal{E_N}(\rho')$, the logarithmic  negativity of $\rho'$ defined as the logarithm of the trace norm of $\tilde \rho'$:
\BE
\mathcal{E_N}(\rho')=\ln || \tilde \rho' ||_1=\max( 0, -\ln \tilde \nu'_-), \label{logneg}
\EE
where $\tilde \nu'_-$ is obtained exlicitly from (\ref{nu}) and  (\ref{UV}), 
\begin{widetext}
 \BE
 \tilde \nu'_-= \frac{1}{ \gamma^2}\left( \sqrt{1-\frac{\delta^2}{N \cosh^2\left(\sqrt{N} \gamma r\right)}} \cosh\left(\sqrt{N} \gamma r\right) - \sinh\left(\sqrt{N} \gamma r\right) \right)^2 . \label {logneg1}
 \EE
 \end{widetext}

We first consider the case of zero phase mismatch,
$\delta=0$, for which (\ref{logneg}) reduces to
\BEA
\tilde \nu'_-&=& \left( \cosh(\sqrt{N} r)  -  \sinh(\sqrt{N}  r)\right)^2\NN\\
&=&e^{-2\sqrt{N}r}.
\EEA
The logarithmic negativity is thus positive and, furthermore, scales quadratically with the number of modes,
 \BE
    \mathcal{E_N}(\rho')=2\sqrt{N}r. \label{lnN}
 \EE
 Note that the covariance matrix  (\ref{CM4}) is block-diagonal since  (\ref{abc}) entails that 
$a=\cosh(2\sqrt{N}r)$, $b=\sinh(2\sqrt{N}r)$ and $c=0$.
It is readily recognized that its nontrivial part pertains to two
entangled modes with a squeezing parameter $2\sqrt{N}r$ and the remainder to
$2N-1$ modes in the vacuum state.
This is one instance of the localization of
entanglement as introduced  in Ref.~\cite{ASI05}.
The result (\ref{lnN}) generalizes the one obtained in Ref.~\cite{DBCK10} for a single pair  of tilted pump ($N=1$) and shows that the effective squeezing  is enhanced by a factor $\sqrt{N}$.

When the phase mismatch takes on some finite values, the logarithmic negativity (\ref{logneg})  features the smallest symplectic eigenvalue (\ref{nu}).
It is instructive to obtain a  more explicit expression for a small phase mismatch by expanding around the case $\delta=0$,
\BE
\mathcal{E_N}(\rho')=2\sqrt{N}r-
\frac{\delta^2 r }{\sqrt{N}} \left(1- \frac{\tanh{\sqrt{N}r}}{\sqrt{N}r}\right)   +O(\delta^4)  . \label{lognegD2}
\EE
The correction is of second order in $\delta$ and  negative. It is a bounded function of $N$ whose magnitude decreases for large $N$.
The positivity of the logarithmic negativity implies that the central mode is entangled with the uniform superposition of the tilted modes. This entanglement localization only slowly decreases when the phase mismath $\delta$ increases. 
This confers some robustness to the entanglement localization process and is somehow in contrast with the genuine multipartite entanglement (\ref{C}) which is more sensitive to the phase mismatch.  
Recalling that the squeezing parameter $r=\sqrt{2} \alpha \lambda \ell$, equation
 (\ref{lognegD2}) also quantifies explicitly the entanglement localization in terms of the pump amplitude $\alpha$, the coupling parameter $\lambda$, the crystal length $\ell$ and provides the scaling with $N$.

\section{Conclusions}

We have presented a simple  active optical scheme for the generation of spatial multipartite entanglement.
It consists in using $2N$ symmetrically tilted plane waves as pump modes in the spatially structured parametric down-conversion process taking place in a nonlinear crystal. We have found the analytical solution of the corresponding model in the rotating wave approximation for arbitrary $N$ and possibly nonzero phase mismatch. We have studied quantitatively the entanglement of the $2N+1$ coupled modes obtained at the output of the crystal. 
When the phase matching condition is satisfied, the system exhibits genuine multipartite entanglement. It has also been shown to subsist for nonzero, albeit small, values of the phase mismatch.

In addition, our scheme provides a realistic proposal for the experimental realization of entanglement localization.
By mixing different spatial modes with beam splitters, we can localize the
entanglement distributed initially among all the $2N+1$ spatial modes
in only two well-defined modes, formed by the linear combinations of the
initial ones. Interestingly, this  entanglement localization results in an enhancement
of the entanglement by a factor $\sqrt{N}$. Moreover, this process has been shown to be robust with respect to the phase mismatch.

\section{Acknowledgments}
The authors are grateful to M. I. Kolobov for stimulating and helpful discussions.
This work was supported by the FET programme COMPAS FP7-ICT-212008.

\end{document}